\documentclass{article}

\usepackage{microtype}
\usepackage{graphicx}
\usepackage{subfigure}
\usepackage{booktabs} 

\usepackage{hyperref}

\usepackage[accepted]{icml2023}

\usepackage{amsmath}
\usepackage{amssymb}
\usepackage{mathtools}
\usepackage{amsthm}

\usepackage{amsmath,amsfonts,bm}

\def\eqref#1{equation~\ref{#1}}

\def\1{\bm{1}}

\newcommand{\beps}{\boldsymbol\epsilon}

\def\rvx{{\mathbf{x}}}

\def\rvz{{\mathbf{z}}}

\def\rmI{{\mathbf{I}}}

\def\vzero{{\bm{0}}}

\def\mI{{\bm{I}}}

\DeclareMathAlphabet{\mathsfit}{\encodingdefault}{\sfdefault}{m}{sl}
\SetMathAlphabet{\mathsfit}{bold}{\encodingdefault}{\sfdefault}{bx}{n}

\def\gL{{\mathcal{L}}}

\def\gN{{\mathcal{N}}}

\usepackage{multirow}
\usepackage[capitalize,noabbrev]{cleveref}

\theoremstyle{plain}

\theoremstyle{definition}

\theoremstyle{remark}

\usepackage[textsize=tiny]{todonotes}

\newenvironment{shrinkeq}[1]
{ \bgroup
\addtolength\abovedisplayshortskip{#1}
\addtolength\abovedisplayskip{#1}
\addtolength\belowdisplayshortskip{#1}
\addtolength\belowdisplayskip{#1}}
{\egroup\ignorespacesafterend}
\icmltitlerunning{Make-An-Audio: Text-To-Audio Generation with Prompt-Enhanced Diffusion Models}

\begin{document}

\twocolumn[
\icmltitle{Make-An-Audio: Text-To-Audio Generation with Prompt-Enhanced Diffusion Models}



\icmlsetsymbol{equal}{*}

\begin{icmlauthorlist}
    \icmlauthor{Rongjie Huang}{equal,zju}
    \icmlauthor{Jiawei Huang}{equal,zju}
    \icmlauthor{Dongchao Yang}{equal,pku}
    \icmlauthor{Yi Ren}{byte}
    \icmlauthor{Luping liu}{zju}
     \icmlauthor{Mingze Li}{zju}
    \icmlauthor{Zhenhui Ye}{zju}
    \icmlauthor{Jinglin Liu}{zju}
    \icmlauthor{Xiang Yin}{byte}
    \icmlauthor{Zhou Zhao}{zju}
    \end{icmlauthorlist}
    
    \icmlaffiliation{zju}{Zhejiang University}
    \icmlaffiliation{pku}{Peking University}
    \icmlaffiliation{byte}{Speech \& Audio Team, ByteDance AI Lab}
    
    \icmlcorrespondingauthor{Zhou Zhao}{ZhaoZhou@zju.edu.cn}

\icmlkeywords{Machine Learning, ICML}

\vskip 0.3in
]



\printAffiliationsAndNotice{\icmlEqualContribution} 

\begin{abstract}
Large-scale multimodal generative modeling has created milestones in text-to-image and text-to-video generation. Its application to audio still lags behind for two main reasons: the lack of large-scale datasets with high-quality text-audio pairs, and the complexity of modeling long continuous audio data. In this work, we propose Make-An-Audio with a prompt-enhanced diffusion model that addresses these gaps by 1) introducing pseudo prompt enhancement with a distill-then-reprogram approach, it alleviates data scarcity with orders of magnitude concept compositions by using language-free audios; 2) leveraging spectrogram autoencoder to predict the self-supervised audio representation instead of waveforms. Together with robust contrastive language-audio pretraining (CLAP) representations, Make-An-Audio achieves state-of-the-art results in both objective and subjective benchmark evaluation. Moreover, we present its controllability and generalization for X-to-Audio with \textbf{``No Modality Left Behind''}, for the first time unlocking the ability to generate high-definition, high-fidelity audios given a user-defined modality input. Audio samples are available at \url{https://Text-to-Audio.github.io}

\end{abstract}

\begin{figure}[h]
    \centering
    \includegraphics[width=0.43\textwidth]{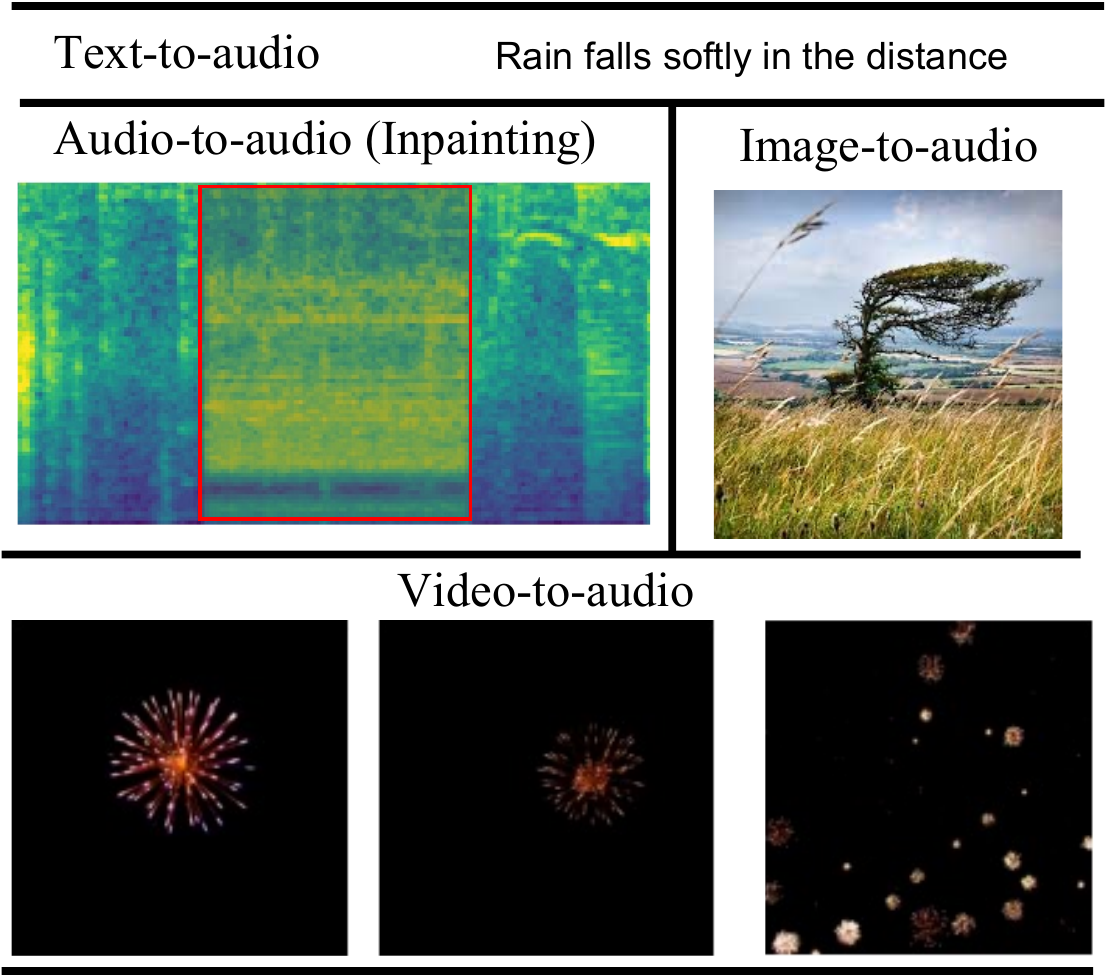}
    \vspace{-3mm}
    \caption{\textbf{No Modality Left Behind:} Make-An-Audio generalizes well to X-to-Audio with multiple user-defined inputs (text, audio, image and video), it empowers humans to create rich and diverse audio content, opening up to a various applications with personalized transfer and fine-grained control.} 
    \label{fig:all_to_audio}
    \vspace{-4mm}
  \end{figure}

\section{Introduction}

Deep generative models~\citep{goodfellow2020generative,kingma2018glow,ho2020denoising} have recently exhibited high-quality samples in various data modalities. With large-scale training data and powerful models, kinds of text-to-image~\citep{saharia2022photorealistic,ramesh2021zero,nichol2021glide} and text-to-video~\citep{singer2022make,hong2022cogvideo} models are now able to vividly depict the visual scene described by a text prompt, and empower humans to create rich and diverse visual content with unprecedented ease. However, replicating this success for audios is limited for the lack of large-scale datasets with high-quality text-audio pairs, and the extreme complexity of modeling long continuous signal data.



In this work, we propose Make-An-Audio, with a prompt-enhanced diffusion model for text-to-audio (T2A) generation. To alleviate the issue of data scarcity, we introduce a pseudo prompt enhancement approach to construct natural languages that align well with audio, opening up the usage of orders of magnitude unsupervised language-free data. To tackle the challenge of modeling complex audio signals in T2A generation, we introduce a spectrogram autoencoder to predict the self-supervised representations instead of waveforms, which guarantees efficient compression and high-level semantic understanding. Together with the power of contrastive language-audio pretraining (CLAP)~\citep{radford2021learning,elizalde2022clap} and high-fidelity diffusion models~\citep{ho2020denoising,song2020denoising,rombach2022high}, it achieves a deep level of language understanding with high-fidelity generation. 


While conceptually simple and easy to train, Make-An-Audio yields surprisingly strong results. Both subjective and objective evaluations demonstrate that Make-An-Audio achieves new state-of-the-art in text-to-audio with natural and controllable synthesis. Make-An-Audio exhibits superior audio quality and text-audio alignment faithfulness on the benchmark AudioCaption dataset and even generalizes well to the unsupervised Clotho dataset in a zero-shot fashion. 


For the first time, we contextualize the need for audio generation with different input modalities. Besides natural language, Make-An-Audio generalizes well to multiple user-defined input modalities (audio, image, and video), which empowers humans to create rich and diverse audio content and opens up a host of applications for personalized transfer and fine-grained control.

Key contributions of the paper include:
\begin{itemize}
  \item We present Make-An-Audio – an effective method that leverages latent diffusion with a spectrogram autoencoder to model the long continuous waveforms.
  \item We introduce a pseudo prompt enhancement with the distill-then-reprogram approach, it includes a large number of concept compositions by opening up the usage of language-free audios to alleviate data scarcity.
  \item We investigate textual representation and emphasize the advantages of contrastive language-audio pretraining for a deep understanding of natural languages with computational efficiency.
  \item We evaluate Make-An-Audio and present state-of-the-art quantitative results and thorough evaluation with qualitative findings.
  \item We generalize the powerful model to X-to-Audio generation, for the first time unlocking the ability to generate high-definition, high-fidelity audios given a user-defined modality input.
\end{itemize}

\section{Related Works}

\subsection{Text-Guided Image/Video Synthesis}

With the rapid development of deep generative models, text-guided synthesis has been widely studied in images and videos. The pioneering work of DALL-E~\citep{ramesh2021zero} encodes images into discrete latent tokens using VQ-VAE~\citep{van2017neural} and considers T2I generation as a sequence-to-sequence translation problem. More recently, impressive visual results have been achieved by leveraging large-scale diffusion models. GLIDE~\citep{nichol2021glide} trains a T2I upsampling model for a cascaded generation. Imagen~\citep{saharia2022photorealistic} presents T2I with an unprecedented degree of photorealism and a deep level of language understanding. Stable diffusion~\citep{rombach2022high} utilizes latent space diffusion instead of pixel space to improve computational efficiency. A large body of work also explores the usage of T2I models for video generation. CogVideo~\citep{hong2022cogvideo} is built on top of a CogView2~\citep{ding2022cogview2} T2I model with a multi-frame-rate hierarchical training strategy. Make-A-Video~\citep{singer2022make} extends a diffusion-based T2I model to T2V through a spatiotemporally factorized diffusion model.

Moving beyond visual generation, our approach aims to generate high-fidelity audio from arbitrary natural language, which has been relatively overlooked. 

\subsection{Text-Guided Audio Synthesis}
While there is remarkable progress in text-guided visual generation, the progress of text-to-audio (T2A) generation lags behind mainly due to two main reasons: the lack of large-scale datasets with high-quality text-audio pairs, and the complexity of modeling long continuous waveforms data. DiffSound~\citep{yang2022diffsound} is the first to explore text-to-audio generation with a discrete diffusion process that operates on audio codes obtained from a VQ-VAE, leveraging masked text generation with CLIP representations. AudioLM~\citep{borsos2022audiolm} introduces the discretized activations of a masked language model pre-trained on audio and generates syntactically plausible speech or music. 

Very recently, the concurrent work AudioGen~\citep{kreuk2022audiogen} propose to generate audio samples autoregressively conditioned on text inputs, while our proposed method differentiates from it in the following: 1) we introduce pseudo prompt enhancement and leverage the power of contrastive language-audio pre-training and diffusion models for high-fidelity generation. 2) We predict the continuous spectrogram representations, significantly improving computational efficiency and reducing training costs.

\begin{figure*}[h]
    \vspace{-2mm}
    \centering
    \includegraphics[width=0.9\textwidth]{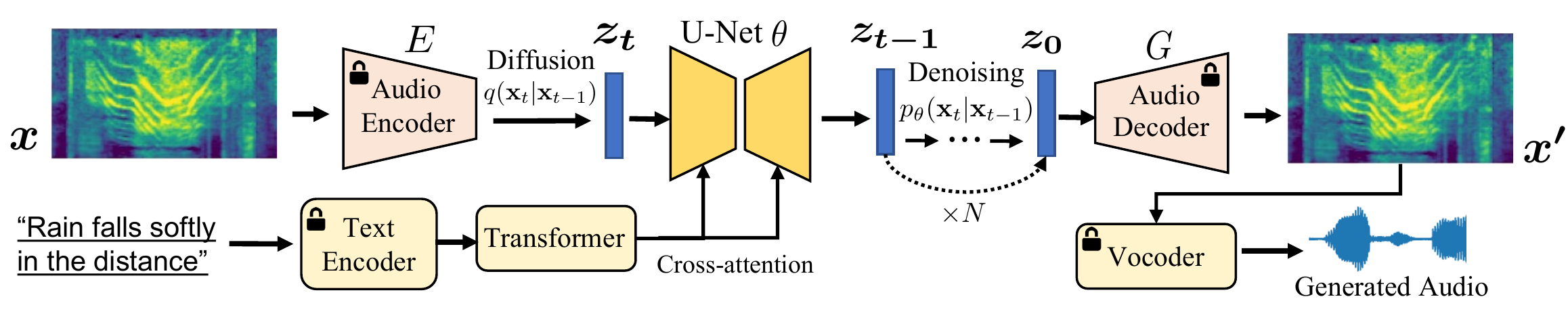}
    \vspace{-4mm}
    \caption{A high-level overview of Make-An-Audio. Note that some modules (printed with a \textit{lock}) are frozen for training the T2A model.} 
    \vspace{-2mm}
    \label{fig:arch}
  \end{figure*}

  \begin{figure*}[h]
    \centering
    \includegraphics[width=0.9\textwidth]{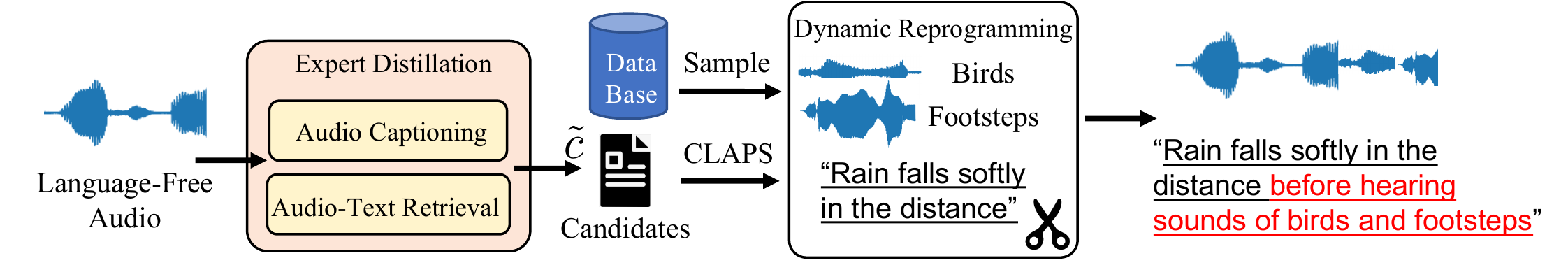}
    \vspace{-4mm}
    \caption{The process of pseudo prompt enhancement. Our semi-parametric diffusion model consists of a fixed expert distillation and a dynamic reprogramming stage. The database $D$ contains audio examples with a sampling strategy $\xi$ to create unseen object compositions. We use CLAPS to denote the CLAP selection.} 
    \vspace{-2mm}
    \label{fig:pseudo_prompt}
  \end{figure*}

\subsection{Audio Representation Learning}
Different from modeling fine-grain details of the signal, the usage of high-level self-supervised learning (SSL)~\citep{baevski2020wav2vec,hsu2021hubert,he2022masked} has been shown to effectively reduce the sampling space of generative algorithms. Inspired by vector quantization (VQ) techniques, SoundStream~\citep{zeghidour2021soundstream} presents a hierarchical architecture for high-level representations that carry semantic information. Data2vec~\citep{baevski2022data2vec} uses a fast convolutional decoder and explores the contextualized target representations in a self-supervised manner. 

Recently, spectrograms (akin to 1-channel 2D images) autoencoder~\citep{gong2022ssast,he2022masked} with reconstruction objective as self-supervision have demonstrated the effectiveness of heterogeneous image-to-audio transfer, advancing the field of speech and audio processing on a variety of downstream tasks. Among these approaches, ~\citet{xu2022masked} study the Masked Autoencoders (MAE)~\citep{he2022masked} to self-supervised representation learning from audio spectrograms. ~\citet{gong2022ssast} adopt audio spectrogram transformer with joint discriminative and generative masked spectrogram modeling. Inspired by these, we inherit the recent success of spectrogram SSL in the frequency domain, which guarantees efficient compression and high-level semantic understanding.

\section{Make-An-Audio}

In this section, we first overview the Make-An-Audio framework and illustrate pseudo prompt enhancement to better align text and audio semantics, following which we introduce textual and audio representations for multimodal learning. Together with the power of diffusion models with classifier-free guidance, Make-An-Audio explicits high-fidelity synthesis with superior generalization.

\subsection{Overview}
Deep generative models have achieved leading performances in text-guided visual synthesis. However, the current development of text-to-audio (T2A) generation is hampered by two major challenges: 1) Model training is faced with data scarcity, as human-labeled audios are expensive to create, and few audio resources provide natural language descriptions. 2) Modeling long continuous waveforms (e.g., typically 16,000 data points for 1s 16 kHz waveforms) poses a challenge for all high-quality neural synthesizers.


As illustrated in Figure~\ref{fig:arch}, Make-An-Audio consists of the following main components: 1) the pseudo prompt enhancement to alleviate the issue of data scarcity, opening up the usage of orders of magnitude language-free audios; 2) a spectrogram autoencoder for predicting self-supervised representation instead of long continuous waveforms; 3) a diffusion model that maps natural language to latent representations with the power of contrastive language-audio pretraining (CLAP) and 4) a separately-trained neural vocoder to convert mel-spectrograms to raw waveforms. In the following sections, we describe these components in detail.

\subsection{Pseudo Prompt Enhancement: Distill-then-Reprogram}

To mitigate the data scarcity, we propose to construct prompts aligned well with audios, enabling a better understanding of the text-audio dynamics from orders of magnitude unsupervised data. As illustrated in Figure~\ref{fig:pseudo_prompt}, it consists of two stages: an expert distillation approach to produce prompts aligned with audio, and a dynamic reprogramming procedure to construct a variety of concept compositions. 

\subsubsection{Expert Distillation}
 
We consider the pre-trained automatic audio captioning~\citep{xu2020crnn} and audio-text retrieval~\citep{deshmukh2022audio,koepke2022audio} systems as our experts for prompt generation. Captioning models aim to generate diverse natural language sentences to describe the content of audio clips. Audio-text retrieval takes a natural language as a query to retrieve relevant audio files in a database. To this end, experts jointly distill knowledge to construct a caption aligned with audio, following which we select from these candidates that endow high CLAP~\citep{elizalde2022clap} score as the final caption (we include a threshold to selectly consider faithful results). This simple yet effective procedure largely alleviates data scarcity issues and explicit generalization to different audio domains, and we refer the reader to Section~\ref{ablation:pseudo} for a summary of our findings. Details have been attached in Appendix~\ref{detail_t2a}.

\subsubsection{Dynamic Reprogramming}
To prevent overfitting and enable a better understanding of concept compositions, we introduce a dynamic reprogramming technique that constructs a variety of concept compositions. It proceeds in three steps as illustrated in Figure~\ref{fig:pseudo_prompt}, where we elaborate the process as follows: 1) We first prepare our sound event database $D$ annotated with a single label. 2) Each time $N$ concepts are sampled from the database $D$, where $N \in \{0, 1, 2\}$. 3) The original text-audio pair data has been randomly concatenated with the sampled events according to the template, constructing a new training example with varied concept compositions. It can be conducted online, significantly reducing the time consumed for data preparation. The reprogramming templates are attached in Appendix~\ref{templates}.

\subsection{Textual Representation}

Text-guided synthesis models need powerful semantic text encoders to capture the meaning of arbitrary natural language inputs, which could be grouped into two major categories: 1) Contrastive pretraining. Similar to CLIP~\citep{radford2021learning} pre-trained on image-text data, recent progress on contrastive language-audio pretraining (CLAP)~\citep{elizalde2022clap} brings audio and text descriptions into a joint space and demonstrates the outperformed zero-shot generalization to multiple downstream domains. 2) Large-scale language modeling (LLM). \citet{saharia2022photorealistic} and \citet{kreuk2022audiogen} utilize language models (e.g., BERT~\citep{devlin2018bert}, T5~\citep{raffel2020exploring}) for text-guided generation. Language models are trained on text-only corpus significantly larger than paired multimodal data, thus being exposed to a rich distribution of text. 

Following the common practice~\citep{saharia2022photorealistic,ramesh2022hierarchical}, we freeze the weights of these text encoders. We find that both CLAP and T5-Large achieve similar results on benchmark evaluation, while CLAP could be more efficient without offline computation of embeddings required by LLM. We refer the reader to Section~\ref{ablation:textual} for a summary of our findings.

\subsection{Audio Representation}
Recently, spectrograms (akin to 1-channel 2D images) autoencoder~\citep{gong2022ssast,he2022masked} with reconstruction objective as self-supervision have demonstrated the effectiveness of heterogeneous image-to-audio transfer, advancing the field of speech and audio processing on a variety of downstream tasks. The audio signal is a sequence of mel-spectrogram sample $\boldsymbol{x} \in[0,1]^{C_{\mathrm{a}} \times T}$, where $C_{\mathrm{a}}, T$ respectively denote the mel channels and the number of frames. Our spectrogram autoencoder is composed of 1) an encoder network $E$ which takes samples $\boldsymbol{x}$ as input and outputs latent representations $z$; 2) a decoder network $G$ reconstructs the mel-spectrogram signals $\boldsymbol{x'}$ from the compressed representation $z$; and 3) a multi-window discriminator $Dis$ learns to distinguish the generated samples $G(z)$ from real ones in different multi-receptive fields of mel-spectrograms.

The whole system is trained end-to-end to minimize 1) Reconstruction loss $\gL_{re}$, which improves the training efficiency and the fidelity of the generated spectrograms; 2) GAN losses $\gL_{GAN}$, where the discriminator and generator play an adversarial game; and 3) KL-penalty loss $\gL_{KL}$, which restricts spectrogram encoders to learn standard $z$ and avoid arbitrarily high-variance latent spaces.

To this end, Make-An-Audio takes advantage of the spectrogram autoencoder to predict the self-supervised representations instead of waveforms. It largely alleviates the challenges of modeling long continuous data and guarantees high-level semantic understanding.

\begin{figure*}[h]
  \centering
  \includegraphics[width=0.7\textwidth]{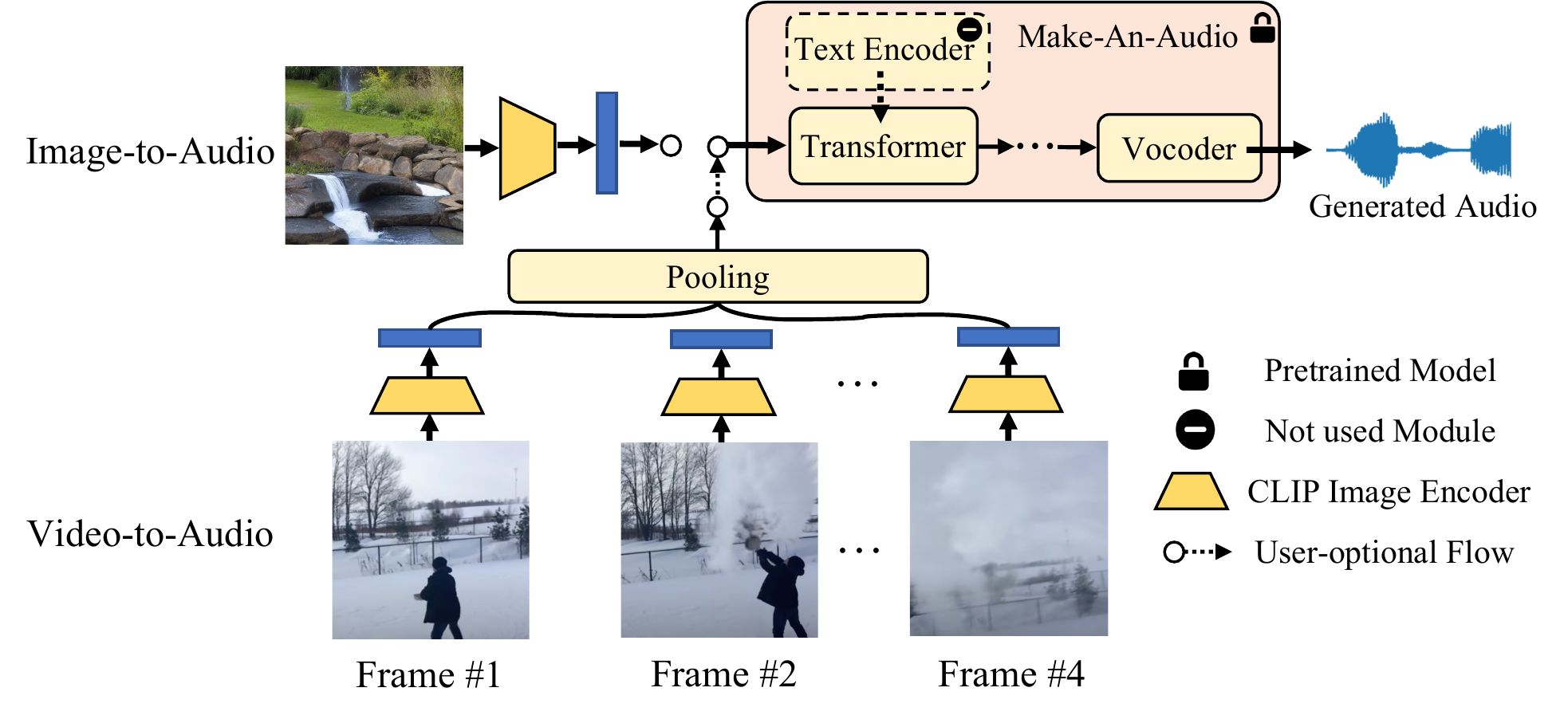}
  \vspace{-4mm}
  \caption{A high-level overview of visual-to-audio generation (I2A/V2A) pipeline using Make-An-Audio.} 
  \vspace{-2mm}
  \label{fig:visual_to_audio}
\end{figure*}

\subsection{Generative Latent Diffusion}

We implement our method over Latent Diffusion Models (LDMs)~\citep{rombach2022high}, a recently introduced class of Denoising Diffusion Probabilistic Models (DDPMs)~\citep{ho2020denoising} that operate in the latent space. It is conditioned on textual representation, breaking the generation process into several conditional diffusion steps. The training loss is defined as the mean squared error in the noise $\beps \sim \mathcal{N}(\vzero,\rmI)$ space, and efficient training is optimizing a random term of $t$ with stochastic gradient descent:
\begin{equation}
    \label{score_loss}
    \gL_{\theta} = \left\lVert \beps_\theta(\rvz_t, t, c)-\beps\right\rVert_2^2,
\end{equation}
where $\alpha$ denotes the small positive constant, and $\beps_\theta$ denotes the denoising network. To conclude, the diffusion model can be efficiently trained by optimizing ELBO without adversarial feedback, ensuring extremely faithful reconstructions that match the ground-truth distribution. Detailed formulation of DDPM has been attached in Appendix~\ref{DDPM}.

\subsection{Classifier-Free Guidance}

For classifier-free guidance shown in~\citep{dhariwal2021diffusion,ho2022classifier}, by jointly training a conditional and an unconditional diffusion model, it could be possible to combine the conditional and unconditional scores to attain a trade-off between sample quality and diversity. The textual condition in a latent diffusion model $\beps_\theta(\rvz_t, t, c)$ is replaced by an empty prompt $c_{\emptyset}$ with a fixed probability during training. During sampling, the output of the model is extrapolated further in the direction of $\beps_\theta(\rvz_t, t, c)$ and away from $\beps_\theta(\rvz_t, t, c_{\emptyset})$ with the guidance scale $s \geq 1$:
\begin{shrinkeq}{-1ex}
\begin{equation}
\small
    \label{guidance}
    \tilde{\beps}_\theta(\rvz_t, t, c) = \beps_\theta(\rvz_t, t, c_{\emptyset}) + s \cdot (\beps_\theta(\rvz_t, t, c) - \beps_\theta(\rvz_t, t, c_{\emptyset}))
\end{equation}
\end{shrinkeq}

\section{X-To-Audio: No Modality Left Behind}

  In this section, we generalize our powerful conditional diffusion model for X-To-Audio generation. For the first time, we contextualize the need for audio generation with different conditional modalities, including: 1) text, 2) audio (inpainting), and 3) visual. Make-An-Audio empowers humans to create rich and diverse audio content with unprecedented ease, unlocking the ability to generate high-definition, high-fidelity audio given a user-defined modality input.



\subsection{Personalized Text-To-Audio Generation}

Adapting models~\citep{chen2021adaspeech,huang2022generspeech} to a specific individual or object is a long-standing goal in machine learning research. More recently, personalization~\citep{gal2022image,benhamdi2017personalized} efforts can be found in vision and graphics, which allows to inject unique objects into new scenes, transform them across different styles, and even produce new products. For instance, when asked to generate ``baby crying'' given the initial sound of ``thunder'', our model produces realistic and faithful audio describing ``a baby cries in the thunder day". Distinctly, it has a wide range of uses for audio mixing and tuning, e.g., adding background sound for an existing clip or editing audio by inserting a speaking object. 

We investigate the personalized text-to-audio generation by stochastic differential editing~\citep{meng2021sdedit}, which has been demonstrated to produce realistic samples with high-fidelity manipulation. Given input audio with a user guide (prompt), we select a particular time $t_0$ with total denoising steps $N$, and add noise to the raw data $\rvz_0$ for $\rvz_T$ ($T = t_0 \times N$) according to Equation~\ref{diffusion_forward}. It is then subsequently denoised through a reverse process parameterized by shared $\theta$ to increase its realism according to Equation~\ref{denoising_backward}.

A trade-off between faithfulness (text-caption alignment) and realism (audio quality) could be witnessed: As $T$ increases, a large amount of noise would be added to the initial audio, and the generated samples become more realistic while less faithful. We refer the reader to Figure~\ref{fig:personalized} for a summary of our findings.



\subsection{Audio Inpainting}
Inpainting~\citep{liu2020rethinking,nazeri2019edgeconnect} is the task of filling masked regions of an audio with new content since parts of the audio are corrupted or undesired. Though diffusion model inpainting can be performed by adding noise to initial audio and sampling with SDEdit, it may result in undesired edge artifacts since there could be an information loss during the sampling process (the model can only see a noised version of the context). To achieve better results, we explicitly fine-tune Make-An-Audio for audio inpainting. 

During training, the way masks are generated greatly influences the final performance of the system. As such, we adopt irregular masks (thick, medium, and thin masks) suggested by LaMa~\citep{suvorov2022resolution}, which uniformly uses polygonal chains dilated by a high random width (wide masks) and rectangles of arbitrary aspect ratios (box masks). In addition, we investigate the frame-based masking strategy commonly adopted in speech liteature~\citep{baevski2020wav2vec,hsu2021hubert}. It is implemented using the algorithm from wav2vec 2.0~\citep{baevski2020wav2vec}, where spans of length are masked with a $p$ probability. 


\subsection{Visual-To-Audio Generation}

Recent advances in deep generative models have shown impressive results in the visually-induced audio generation~\citep{su2020audeo,gan2020foley}, towards generating realistic audio that describes the content of images or videos: ~\citet{hsu2020text} show that spoken language could be learned by a visually-grounded generative model of speech. ~\citet{iashin2021taming} propose a multi-class visual guided sound synthesis that relies on a codebook prior-based transformer.

To pursue this research further, we extend Make-An-Audio for visual-to-audio generation. For the lack of large-scale visual-audio datasets in image-to-audio (I2A) research, our main idea is to utilize contrastive language-image pretraining (CLIP) with CLIP-guided T2A model and leverage textual representations to bridge the modality gap between visual and audio world. As CLIP encoders embed images and text to the joint latent space, our T2A model provides a unique opportunity to visualize what the CLIP image encoder is seeing. Considering the complexity of V2A generation, it is natural to leverage image priors for videos to simplify the learning process. On this account, we uniformly pick up 4 frames from the video and pool these CLIP image features to formulate the ``averaged" video representation, which is then deteriorated to I2A generation.


To conclude, the visual-to-audio inference scheme can be formulated in Figure~\ref{fig:visual_to_audio}. It significantly reduces the requirement for pair visual datasets, and the plug-and-play module with pre-trained Make-An-Audio empowers humans to create rich and diverse audio content from the visual world.
\section{Training and Evaluation}

\subsection{Dataset}
We train on a combination of several datasets: AudioSet, BBC sound effects, Audiostock, AudioCaps-train, ESC-50, FSD50K, Free To Use Sounds, Sonniss Game Effects, WeSoundEffects, MACS, Epidemic Sound, UrbanSound8K, WavText5Ks, LibriSpeech, and Medley-solos-DB. For audios without natural language annotation, we apply the pseudo prompt enhancement to construct captions aligned well with the audio. Overall we have $\sim$3k hours with 1M audio-text pairs for training data. For evaluating text-to-audio models~\citep{yang2022diffsound,kreuk2022audiogen}, the AudioCaption validation set is adopted as the standard benchmark, which contains 494 samples with five human-annotated captions in each audio clip. For a more challenging zero-shot scenario, we also provide results in Clotho~\citep{drossos2020clotho} validation set which contain multiple audio events. A more detailed data setup has been attached in Appendix~\ref{app:data}.

We conduct preprocessing on the text and audio data: 1) convert the sampling rate of audios to 16kHz and pad short clips to 10-second long; 2) extract the spectrogram with the FFT size of 1024, hop size of 256 and crop it to a mel-spectrogram of size $80 \times 624$; 3) non-standard words (e.g., abbreviations, numbers, and currency expressions) and semiotic classes~\citep{taylor2009text} (text tokens that represent particular entities that are semantically constrained, such as measure phrases, addresses, and dates) are normalized.


\subsection{Model Configurations}
We train a continuous autoencoder to compress the perceptual space with downsampling to a 4-channel latent representation, which balances efficiency and perceptually faithful results. For our main experiments, we train a U-Net~\citep{ronneberger2015u} based text-conditional diffusion model, which is optimized using 18 NVIDIA V100 GPU until 2M optimization steps. The base learning rate is set to 0.005, and we scale it by the number of GPUs and the batch size following LDM. We utilize HiFi-GAN~\cite{kong2020hifi} (V1) trained on VGGSound dataset~\citep{chen2020vggsound} as the vocoder to synthesize waveform from the generated mel-spectrogram in all our experiments. Hyperparameters are included in Appendix~\ref{appendix:model}.

\subsection{Evaluation Metrics}
We evaluate models using objective and subjective metrics over audio quality and text-audio alignment faithfulness. Following common practice~\citep{yang2022diffsound,iashin2021taming}, the key automated performance metrics used are melception-based~\citep{koutini2021efficient} FID~\citep{heusel2017gans} and KL divergence to measure audio fidelity. Additionally, we introduce the CLAP score to measure audio-text alignment for this work. CLAP score is adapted from the CLIP score~\citep{hessel2021clipscore,radford2021learning} to the audio domain and is a reference-free evaluation metric that closely correlates with human perception. 

For subjective metrics, we use crowd-sourced human evaluation via Amazon Mechanical Turk, where raters are asked to rate MOS (mean opinion score) on a 20-100 Likert scale. We assess the audio quality and text-audio alignment faithfulness by respectively scoring MOS-Q and MOS-F, which is reported with 95\% confidence intervals (CI). More information on evaluation has been attached in Appendix~\ref{appendix:eval}.


\begin{table*}[ht]
    \centering
    \small
    \begin{tabular}{c|cc|ccc|cc|cc}
    \toprule    
    Model & Text-cond & Params  & FID  & KL  & CLAP  & MOS-Q  & MOS-F   & FID-Z  & KL-Z  \\
    \midrule
    Reference                     & /          & /       &  / &  /   &  0.526                           & 74.7$\pm$0.94   &  80.5$\pm$1.84  & /    & /  \\
    \midrule
    Diffsound                     & CLIP    & 520M    & 7.17   &  3.57 & 0.420                      & 67.1$\pm$1.03   &  70.9$\pm$1.05 & 24.97  & 6.53  \\
    \midrule                 
    \multirow{4}{*}{Make-An-Audio}    & \bf CLAP   & \bf 332M   &\textbf{4.61}  & \textbf{2.79}   &0.482  & \textbf{72.5$\pm$0.90}  & \textbf{78.6$\pm$1.01} & 17.38 & \textbf{6.98} \\
    \cline{2-10}
   \rule{0pt}{10pt}   & BERT    & 809M   &5.15    & 2.89  & 0.480                     & 70.5$\pm$0.87    &  77.2$\pm$0.98 & 18.75 & 7.01 \\
                                    & T5-Large &  563M   &4.83    & 2.81   &\textbf{0.486}          & 71.8$\pm$0.91    &  77.2$\pm$0.93 & \textbf{17.23} & 7.02 \\
                                    & CLIP     &  576M  &6.45    & 2.91   &0.444                    & 72.1$\pm$0.92    &  75.4$\pm$0.96 & 17.55 & 7.09    \\
                                
    \bottomrule
    \end{tabular}
    \vspace{-2mm}
    \caption{Text-to-audio evaluation. We report the evaluation metrics including MOS($\uparrow$), FID($\downarrow$), KL($\downarrow$), and CLAP($\uparrow$). FID-Z and KL-Z denote the zero-shot results in the Clotho dataset.}
    \label{table:text-to-audio}
    \vspace{-6mm}
    \end{table*}

    \begin{table*}[ht]
    \centering
    \small
    \hspace{-0.5cm}
    \begin{minipage}{0.60\textwidth}
        \vspace{6mm}
        \scalebox{0.9}{
        \begin{tabular}{c|ccc|ccc}
    \toprule  
    \multirow{2}{*}{Training Masks} & \multicolumn{3}{c|}{Narrow Masks} & \multicolumn{3}{c}{Wide Masks} \\
                                      & FID     & KL    & MOS-Q    & FID            & KL   & MOS-Q          \\
    \midrule
    Irregular (Thin)                  & 1.83     &0.46 & 68.3$\pm$1.38 & 4.01       & 0.86    &  66.2$\pm$1.20   \\
    Irregular (Medium)                & 1.76     &0.31 & 67.8$\pm$1.41 & 3.93       & 0.65    &  66.9$\pm$1.22   \\
    Irregular (Thick)                 & 1.73     &0.32 & 69.6$\pm$1.36 & 3.83       & 0.67    &  69.3$\pm$1.05   \\
    \midrule
    Frame (p=30\%)                   &  1.64	 &0.29 & 66.9$\pm$1.60 & 3.68       & 0.62    & 66.1$\pm$1.29  \\
    Frame (p=50\%)                   &  1.77     &0.32 & 68.6$\pm$1.42 & 3.66       & 0.63    & 67.4$\pm$1.27  \\
    Frame (p=70\%)                   &  1.59     &0.32 & 71.0$\pm$1.12 & 3.49       & 0.65    & 70.8$\pm$1.50  \\
    \bottomrule
    \end{tabular}}
    \vspace{-2mm}
    \caption{Audio inpainting evaluation with variety masking strategies.}
    \label{table:audio-inpainting}
    \end{minipage}
    \hspace{-0.5cm}
    \vspace{-2cm}
    \begin{minipage}[t]{0.32\linewidth}
    \centering
    \small
    \scalebox{0.9}{
    \begin{tabular}{c|cc}
    \toprule  
    Method                      & MOS-Q        & MOS-F          \\
    \midrule
    \multicolumn{3}{l}{\textbf{Image-to-Audio Generation}} \\
    \midrule
    Reference                  & 72.0$\pm$1.54 &    76.4$\pm$1.83  \\
    Make-An-Audio              & 68.4$\pm$1.09 &    78.0$\pm$1.20   \\
    \midrule  
    \multicolumn{3}{l}{\textbf{Video-to-Audio Generation}} \\
    \midrule  
    Reference                  & 69.5$\pm$1.22 &    81.0$\pm$1.43  \\
    Make-An-Audio              & 60.0$\pm$1.31 &    69.0$\pm$1.08   \\
    \bottomrule
    \end{tabular}}
    \vspace{-3mm}
    \caption{Image/Video-to-audio evaluation.}
    \label{table:visual2audio}
    \end{minipage}
    \vspace{1.9cm}
    \end{table*}

\section{Results}

\subsection{Quantitative Results}

\textbf{Automatic Objective Evaluation} 
The objective evaluation comparison with baseline Diffsound (the only publicly-available T2A generation model) are presented in Table~\ref{table:text-to-audio}, and we have the following observations: 1) In terms of audio qualty, Make-An-Audio achieves the highest perceptual quality in AudioCaption with FID of 4.61 and KL of 2.79. For zero-shot generation, it also demonstrates the outperformed results superior to the baseline model; 2) On text-audio similarity, Make-An-Audio scores the highest CLAP with a gap of 0.037 compared to the ground truth audio, suggesting Make-An-Audio's ability to generate faithful audio that aligns well with descriptions.

\textbf{Subjective Human Evaluation} The evaluation of the T2A models is very challenging due to its subjective nature in perceptual quality, and thus we include a human evaluation in Table~\ref{table:text-to-audio}: Make-An-Audio (CLAP) achieves the highest perceptual quality with MOS-Q of 72.5 and MOS-F of 78.6. It indicates that raters prefer our model synthesis against baselines in terms of audio naturalness and faithfulness. 


For audio-inpainting, we compare different masking designs, including the irregular (thick, medium, and thin) strategy from visual world~\citep{suvorov2022resolution}, as well as the frame-based (with varying $p$) strategy commonly used in speech~\citep{baevski2020wav2vec,hsu2021hubert}. During evaluation, we randomly mask the \textit{wide} or \textit{narrow} regions and utilize FID and KL metrics to measure performance. The results have been presented in Table~\ref{table:audio-inpainting}, and we have the following observations: 1) In both frame-based or irregular strategies, larger masked regions in training have witnessed the improved perceptual quality, which force the network to exploit the high receptive field of continuous spectrograms fully. 2) With the similar size of the masked region, the frame-based strategy consistently outperforms the irregular one, suggesting that it could be better to mask the audio spectrograms which align in time series.

We also present our visual-to-audio generation results in Table~\ref{table:visual2audio}. As can be seen, Make-An-Audio can generalize to a wide variety of images and videos. Leveraging contrastive pre-training, the model provides a high-level understanding of visual input, which generates high-fidelity audio spectrograms well-aligned with their semantic meanings. 

\begin{figure*}[h]
    \centering
    \includegraphics[width=1.0\textwidth]{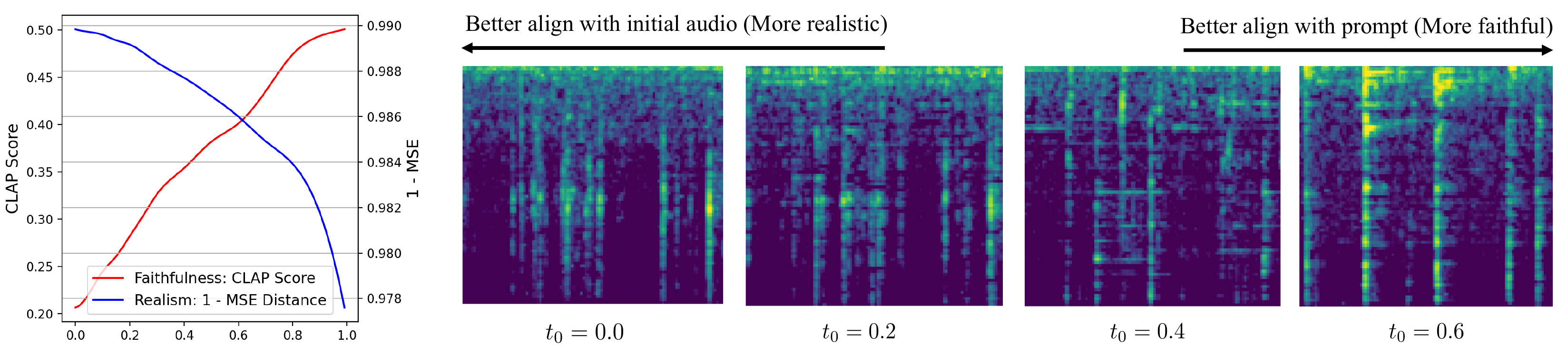}
    \vspace{-6mm}
    \caption{We illustrate personalized text-to-audio results with various $t_0$ initializations. $t_0 = 0$ indicates the initial audio itself, whereas $t_0 = 1$ indicates a text-to-audio synthesis from scratch. For comparison, realism is measured by the 1-MSE distance between generated and initial audio, and faithfulness is measured by the CLAP score between the generated sample. Prompt: A clock ticktocks.}
    \vspace{-2mm}
 \label{fig:personalized}
  \end{figure*}

\subsection{Qualitative Findings}

\begin{figure}[h]
    \centering
    \includegraphics[width=0.5\textwidth]{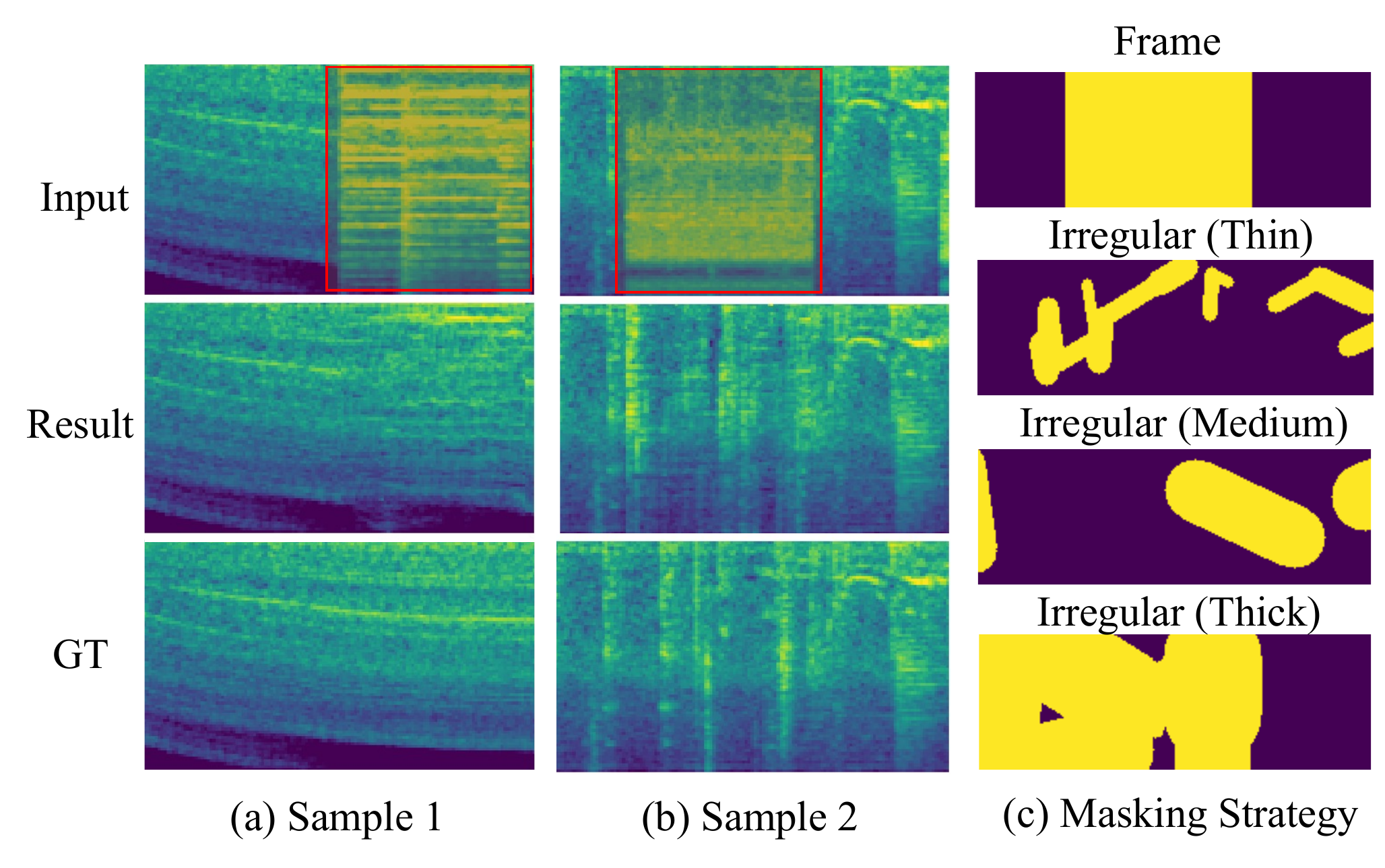}
    \vspace{-7mm}
    \caption{Qualitative results with our inpainting model.} 
    \label{fig:inpaint}
  \end{figure}

  \begin{figure}[h]
    \centering
    \includegraphics[width=0.35\textwidth]{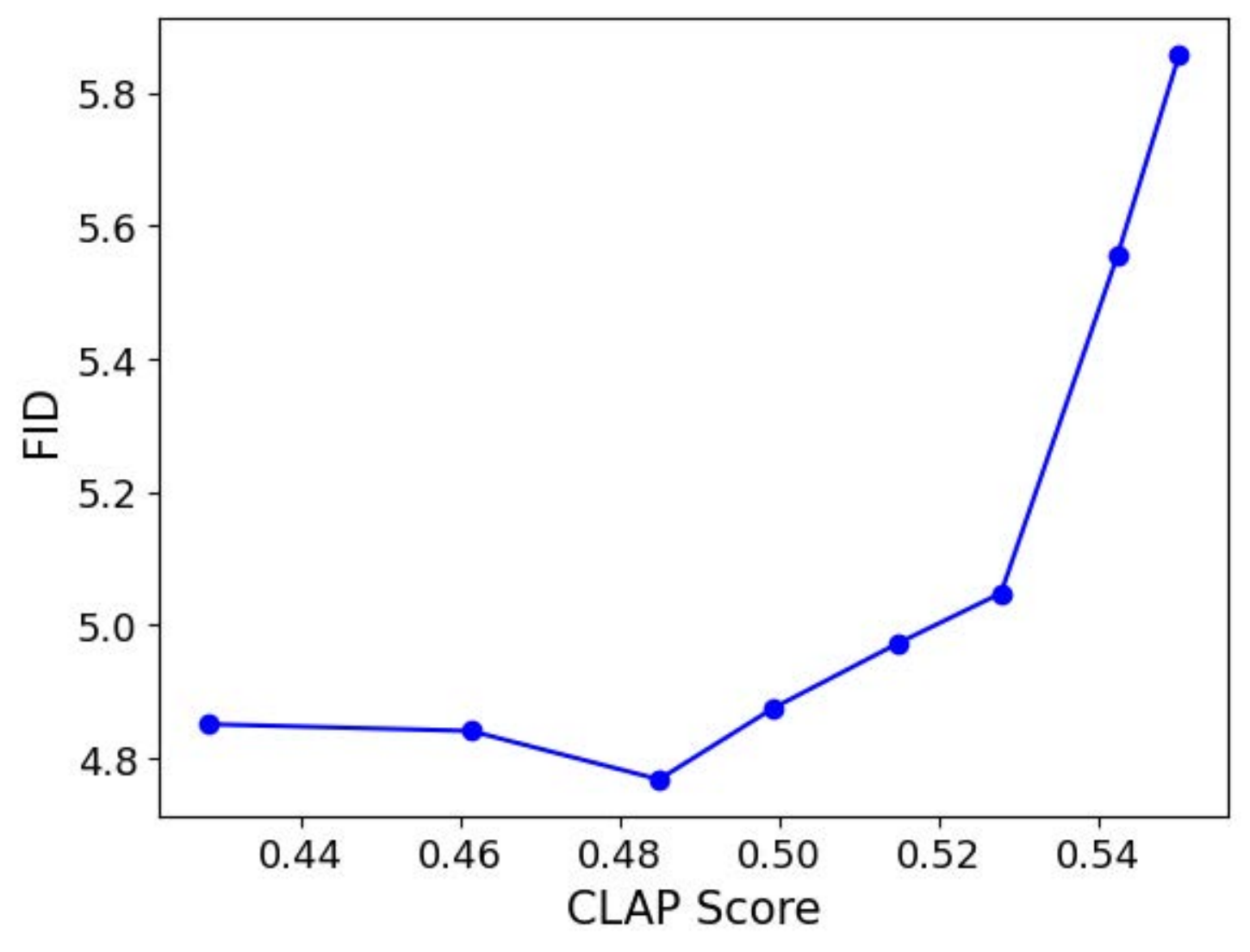}
    \vspace{-2mm}
    \caption{Classifier-free guidance trade-off curves.} 
    \vspace{-3mm}
    \label{fig:CLAP-FID}
  \end{figure}

Firstly, we explore the classifier-free guidance in text-to-audio synthesis. We sweep over guidance values and present trade-off curves between CLAP and FID scores in Figure~\ref{fig:CLAP-FID}. Consistent with the observations in ~\citet{ho2022classifier}, the choice of the classifier guidance weight could scale conditional and unconditional synthesis, offering a trade-off between sample faithfulness and realism with respect to the conditioning text. 

For better comparison in audio inpainting, we visualize different masking strategies and synthesis results in Figure~\ref{fig:inpaint}. As can be seen, given the initial audio with undesired content, our model correctly fills and reconstruct the audio robust to different shapes of masked regions, suggesting that it is capable of a high-level understanding of audio content.

On the personalized text-to-audio generation, we explore different $t_0 \in(0,1)$ to add Gaussian noise and conduct reverse sampling. As shown in Figure~\ref{fig:personalized}, a trade-off between faithfulness (measured by CLAP score) and realism (measured by 1-MSE distance) could be witnessed. We find that $t_0 \in[0.2, 0.5]$ works well for faithful guidance with realistic generation, suggesting that audio variants (e.g., speed, timbre, and energy) could be easily destroyed as $t_0$ increases.


\subsection{Analysis and Ablation Studies}

To verify the effectiveness of several designs in Make-An-Audio, including pseudo prompt enhancement, textual and audio representation, we conduct ablation studies and discuss the key findings as follows. More analysis on audio representation has been attached in Appendix~\ref{autoencoder}.

\subsubsection{Textual Representation} \label{ablation:textual}
We explore several pretrained text encoders, including language models BERT~\citep{devlin2018bert}, T5-Large~\citep{raffel2020exploring}, as well as the multimodal contrastive pre-trained encoder CLIP~\citep{radford2021learning} and CLAP~\citep{elizalde2022clap}. We freeze the weights of text encoders for T2A generation. For easy comparison, we present the results in Table~\ref{table:text-to-audio} and have the following observations: 1) Since CLIP is introduced as a scalable approach for learning joint representations between text and images, it could be less useful in deriving semantic representation for T2A in contrast to~\citet{yang2022diffsound}. 2) CLAP and T5-Large achieve similar performances on benchmarks dataset, while CLAP could be more computationally efficient (with only \%59 params), without the need for offline computation of embeddings in large-scale language models.

\subsubsection{Pseudo Prompt Enhancement} \label{ablation:pseudo}
Our prompt enhancement approach alleviates the issue of data scarcity, which consists of two stages with a distill-then-reprogram approach. As shown in Table~\ref{table:dataset} in Appendix~\ref{app:data}, we calculate and compare the prompt-audio faithfulness averaged across datasets: The joint expert distillation produces high-quality captions aligned well with audio, and suggests strong generalization to diverse audio domains.

To highlight the effectiveness of the proposed dynamic reprogramming strategy to create unseen object compositions, we additionally train our Make-An-Audio in the static training dataset, and attach the results in Table~\ref{table:ablation} in Appendix~\ref{app:implementation}: 1) Removing the dynamic reprogramming approach results in a slight drop in evaluation; 2) When migrating to a more challenging scenario to Clotho in a zero-shot fashion, a significant degradation could be witnessed, demonstrating its effectiveness in constructing diverse object compositions for better generalization.

\section{Conclusion}
In this work, we presented Make-An-Audio with a prompt-enhanced diffusion model for text-to-audio generation. Leveraging the prompt enhancement with the distill-then-reprogram approach, Make-An-Audio was endowed with various concept compositions with orders of magnitude unsupervised data. We investigated textual representation and emphasized the advantages of contrastive pre-training for a deep understanding of natural languages with computational efficiency. Both objective and subjective evaluation demonstrated that Make-An-Audio achieved new state-of-the-art results in text-to-audio with realistic and faithful synthesis. Make-An-Audio was the first attempt to generate high-definition, high-fidelity audio given a user-defined modality input, opening up a host of applications for personalized transfer and fine-grained control. We envisage that our work serve as a basis for future audio synthesis studies.

\clearpage

\bibliography{main}
\bibliographystyle{icml2023}

\appendix
\clearpage
\onecolumn

\begin{center}{\bf {\LARGE Appendices} }
\end{center}
\begin{center}{\bf {\Large Make-An-Audio: Text-To-Audio Generation with Prompt-Enhanced Diffusion Models} \linebreak}
\end{center}

\section{Detailed Experimental Setup}  \label{app:data}

\begin{table}[ht]
    \centering
    \small
    \begin{tabular}{llll}
    \toprule
    Dataset & Hours & Type & Source \\
    \midrule
    Clotho               &  152     &  Caption    &      \citet{drossos2020clotho}   \\   
    AudioCaps            &  109     &  Caption    &   \citet{kim2019audiocaps}     \\
    MACS                 &   100    &  Caption    &    \citet{martin2021ground}    \\
    WavText5Ks           &   25    &   Caption   &     \citet{deshmukh2022audio}  \\  
    BBC sound effects    &  481     &  Caption  &  \url{https://sound-effects.bbcrewind.co.uk/}    \\
    Audiostock           &   43    &  Caption  &     \url{https://audiostock.net/se}    \\ 
    \midrule
    Filter AudioSet      &  2084     &  Label    &   \citet{gemmeke2017audio}     \\
    ESC-50               &  3     &  Label    &     \citet{piczak2015esc}    \\   
    FSD50K               &  108     &  Label     &    \url{https://annotator.freesound.org/fsd/}  \\
    Sonniss Game Effects &  20     &  Label &     \url{https://sonniss.com/gameaudiogdc/}  \\
    WeSoundEffects       &  11     &  Label &   \url{https://wesoundeffects.com/}   \\   
    Epidemic Sound       &  220  &  Label &   \url{https://www.epidemicsound.com/}    \\   
    UrbanSound8K         &  8    &  Label &     \citet{Salamon:UrbanSound:ACMMM:14}   \\
    \midrule
    LibriTTS            &   300   & Language-free    &  \citet{zen2019libritts}  \\
    Medley-solos-DB      &  7     & Language-free    &     \citet{bittner2014medleydb}  \\   
    \bottomrule
    \end{tabular}
    \label{dataset}
    \caption{Statistics for the combination of several datasets.}
    \end{table}

As shown in Table~\ref{table:dataset}, we collect a large-scale audio-text dataset consisting of 1M audio samples with a total duration of $\sim$3k hours. It contains audio of human activities, natural sounds, and audio effects, consisting of several data sources from publicly available websites. For audio with text descriptions, we download the parallel audio-text data. For audios without natural language annotation (or with labels), we discard the corresponding class label (if any) and apply the pseudo prompt enhancement to construct natural language descriptions aligned well with the audio. 

As speech and music are the dominant classes in Audioset, we filter these samples to construct a more balanced dataset. Overall we are left with ~3k hours with 1M audio-text pairs for training data. For evaluating text-to-audio models~\citep{yang2022diffsound,kreuk2022audiogen}, the AudioCaption validation set is the standard benchmark, which contains 494 samples with five human-annotated captions in each audio clip. In both training and inference, we pad short clips to 10-second long and randomly crop a $624 \times 80$ mel-spectrogram from 10-second 16 kHz audio.

\begin{table}[ht]
    \centering
    \small
    \vspace{2mm}
    \begin{tabular}{c|ccc}
        \toprule  
    Method                        & FSD50K            & ESC-50  & Urbansound8k \\
    \midrule
    Original               & 0.40          & 0.43   &  0.33 \\
    \midrule
    Captioning             & 0.35          & 0.46   &  0.37 \\
    Retrieval              & 0.31          & 0.44   &  0.38  \\
     \midrule
    Both + CLAP Select            & 0.54          & 0.62   &  0.55    \\
    \bottomrule                                               
    \end{tabular}   
    \caption{Text-audio alignment CLAP score averaged across the single-label dataset.}
    \label{table:dataset}
    \end{table}

\section{Model Configurations} \label{appendix:model}
We list the model hyper-parameters of Make-An-Audio in Table~\ref{tab:hyperparameters_ps}.

\begin{table}[h]
\small
\centering
\begin{tabular}{l|c|c}
\toprule
\multicolumn{2}{c|}{Hyperparameter}   & Make-An-Audio \\ 
\midrule
\multirow{5}{*}{Spectrogram Autoencoders} 
&Input/Output Channels                    & 1      \\
&Hidden Channels                          &   4   \\ 
&Residual Blocks                         &   2   \\   
&Spectrogram Size                &  $80 \times 624$ \\    
&Channel Mult                &   $[1, 2, 2, 4]$ \\                       
\midrule
\multirow{6}{*}{Denoising Unet} 
&Input/Output Channels       &  4   \\         
&Model Channels              &  320 \\ 
&Attention Heads             &  8 \\
&Condition Channels          &  1024 \\
&Latent Size                  &  $10 \times 78$\\
&Channel Mult                &   $[1, 2]$ \\        
\midrule
\multirow{3}{*}{CLAP Text Encoder} 
&Transformer Embed Channels        &  768   \\         
&Output Project Channels   &  1024 \\    
&Token Length   &  77 \\         
\midrule
\multicolumn{2}{c|}{Total Number of Parameters}   & 332M  \\
\bottomrule
\end{tabular}
\caption{Hyperparameters of Make-An-Audio models.}
\label{tab:hyperparameters_ps}
\end{table}

\section{Evaluation} \label{appendix:eval}

To probe audio quality, we conduct the MOS (mean opinion score) tests and explicitly instruct the raters to \textit{``focus on examining the audio quality and naturalness.''}. The testers present and rate the samples, and each tester is asked to evaluate the subjective naturalness on a 20-100 Likert scale.

To probe text-audio alignment, human raters are shown an audio and a prompt and asked \textit{``Does the natural language description align with audio faithfully?''}. They must respond with ``completely'', ``mostly'', or ``somewhat'' on a 20-100 Likert scale. 

\begin{figure*}[!h]
	\centering
    \subfigure[Screenshot of MOS-F testing.]
    {
    \includegraphics[width=0.95\textwidth]{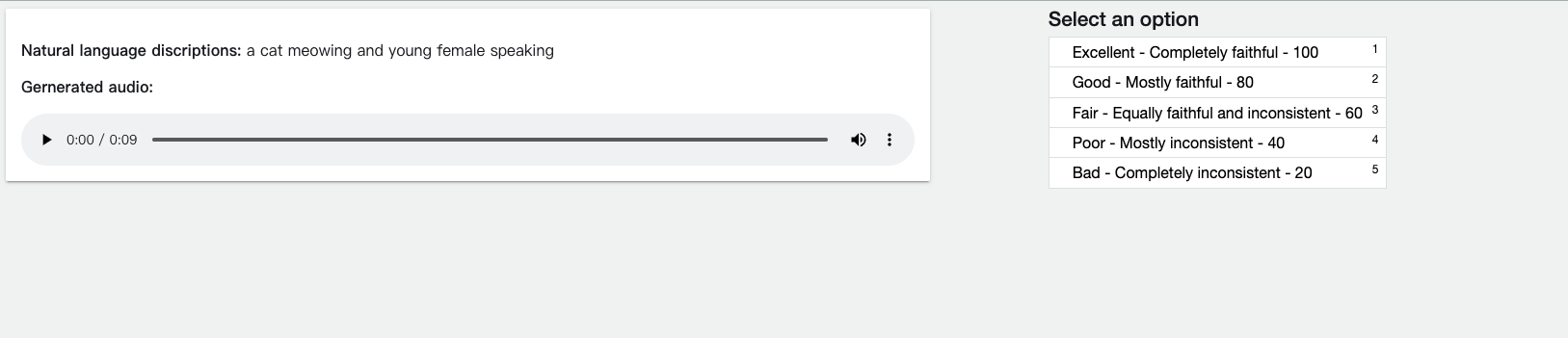}
    }
    \subfigure[Screenshot of MOS-Q testing.]
    {
    \includegraphics[width=0.95\textwidth]{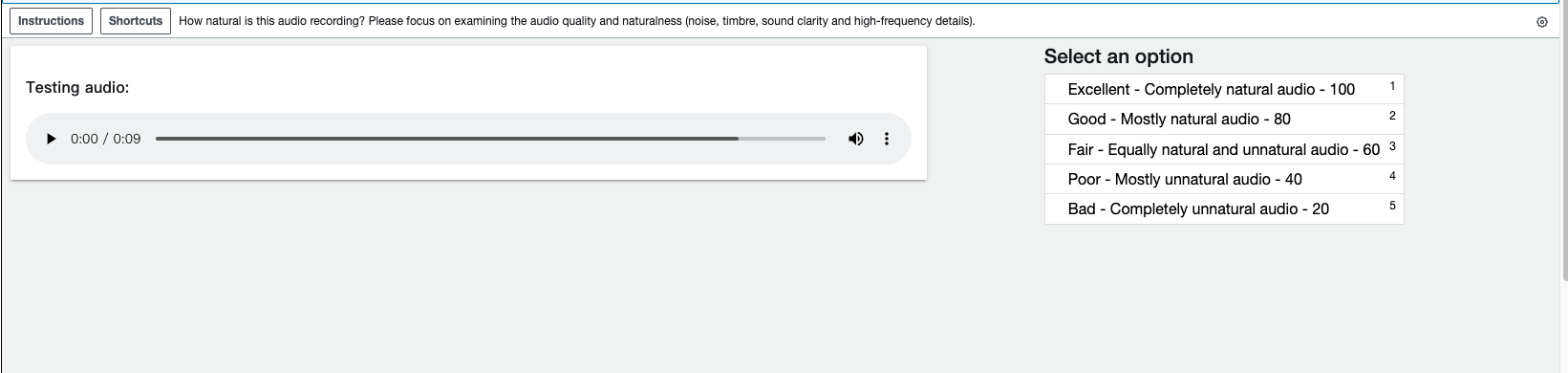}
    }
	\caption{Screenshots of subjective evaluations.}
	\label{fig:screenshot_eval}
\end{figure*}

Our subjective evaluation tests are crowd-sourced and conducted via Amazon Mechanical Turk. These ratings are obtained independently for model samples and reference audio, and both are reported. The screenshots of instructions for testers have been shown in Figure~\ref{fig:screenshot_eval}. We paid \$8 to participants hourly and totally spent about \$750 on participant compensation. A small subset of speech samples used in the test is available at \url{https://Text-to-Audio.github.io/}.

\section{Detailed Formulation of DDPM} \label{DDPM}

We define the data distribution as $q(\rvx_{0})$. The diffusion process is defined by a fixed Markov chain from data $\rvx_0$ to the latent variable $\rvx_T$:
\begin{equation} \label{q_func} 
q(\rvx_{1},\cdots,\rvx_T|\rvx_0) = \prod_{t=1}^T q(\rvx_t|\rvx_{t-1}), 
\quad\ \ 
\end{equation}
For a small positive constant $\beta_t$, a small Gaussian noise is added from $\rvx_{t-1}$ to the distribution of $\rvx_{t}$ under the function of $q(\rvx_t|\rvx_{t-1})$.

The whole process gradually converts data $\rvx_0$ to whitened latents $\rvx_T$ according to the fixed noise schedule $\beta_1,\cdots,\beta_T$, where $\beps\sim\gN(\vzero, \mI)$:
\begin{equation}   \label{diffusion_forward} 
q(\rvx_t|\rvx_{t-1}) := \gN(\rvx_t;\sqrt{1-\beta_t}\rvx_{t-1},\beta_t \mI)
\end{equation}
Efficient training is optimizing a random term of $t$ with stochastic gradient descent: 
\begin{equation}
    \label{eq: score_loss}
    \gL_{\theta} = \left\lVert \beps_\theta\left(\alpha_t\rvx_{0}+\sqrt{1-\alpha_t^2}\beps\right)-\beps\right\rVert_2^2
\end{equation}
Unlike the diffusion process, the reverse process is to recover samples from Gaussian noises. The reverse process is a Markov chain from $x_T$ to $x_0$ parameterized by shared $\theta$:
\begin{equation} \label{denoising_backward}
p_{\theta}(\rvx_0,\cdots,\rvx_{T-1}|\rvx_T)=\prod_{t=1}^T p_{\theta}(\rvx_{t-1}|\rvx_t),
\end{equation}
where each iteration eliminates the Gaussian noise added in the diffusion process:
\begin{equation}
    p_{\theta}(\rvx_{t-1}|\rvx_{t}) := \mathcal{N}(\rvx_{t-1};\mu_{\theta}(\rvx_t,t), \sigma_{\theta}(\rvx_t,t)^2\mI)
\end{equation}

\section{Implementation Details}  \label{app:implementation}

\subsection{Spectrogram Autoencoders} \label{autoencoder}

We also investigate the effectiveness of several audio autoencoder variants in Table~\ref{table:ablation}, and find that deeper representation (i.e., 32 or 128) relatively brings more compression, while the information deterioration could burden the Unet model in generative modeling.


\begin{table}[ht]
    \centering
    \small
    \vspace{2mm}
    \begin{tabular}{c|cccc}
    \toprule  
    Method             & Channel          & FID & KL  \\
    \midrule
    \multicolumn{4}{c}{\bf Supervised Evaluation in AudioCaps dataset}  \\  \midrule
    \multirow{3}{*}{Base} & 4 &  5.15 & 2.89   \\
   & 32 & 9.22  &  3.54 \\ 
     & 128 & 10.92 &  3.68  \\ \midrule
     w/o PPE & 4  & 5.37 &  3.05  \\  \midrule
     \multicolumn{4}{c}{\bf Zero-Shot Evaluation in Clotho dataset}  \\  \midrule
     Base & 4& 18.75 & 7.01 \\
     w/o PPE  & 4  & 22.31 &  7.19 \\
    \bottomrule
    \end{tabular}
    \caption{Audio quality comparisons for ablation study with Make-An-Audio BERT. We use PPE to denote pseudo prompt enhancement.}
    \label{table:ablation}
    \end{table}

\subsection{Text-to-audio} \label{detail_t2a}
We first encode the text into a sequence of K tokens, and utilize the cross-attention mechanism to learn a language and mel-spectrograms representation mapping in a powerful model. After the initial training run, we fine-tuned our base model to support unconditional generation, with 20\% of text token sequences being replaced with the empty sequence. This way, the model retains its ability to generate text-conditional outputs, but can also generate spectrogram representation unconditionally.

We consider the pre-trained automatic audio captioning~\citep{xu2020crnn} and audio-text retrieval~\citep{deshmukh2022audio,koepke2022audio} systems as our experts for prompt generation. Regarding automatic audio captioning, the model consists of a 10-layer convolution neural network (CNN) encoder and a temporal attentional single-layer gated recurrent unit (GRU) decoder. The CNN encoder is pre-trained on a large-scale Audioset dataset. As for audio-text retrieval, the model leverages BERT with a multi-modal transformer encoder for representation learning. It is trained on AudioCaps and Clotho datasets.

\subsection{Visual-to-audio}  \label{detail_v2a}
For visual-to-audio (image/video) synthesis, we utilize the CLIP-guided T2A model and leverage global textual representations to bridge the modality gap between the visual and audio worlds. However, we empirically find that global CLIP conditions have a limited ability to control faithful synthesis with high text-audio similarity. On that account, we use the 110h FSD50K audios annotated with a class label for training, and this simplification avoids multimodal prediction (a conditional vector may refer to different concepts) with complex distribution. 

We conduct ablation studies to compare various training settings, including datasets and global conditions. The results have been presented in Table~\ref{table:ablation_v2a}, and we have the following observations: 1) Replacing the FSD50K dataset with AudioCaps~\citep{kim2019audiocaps} have witnessed a significant decrease in faithfulness. The dynamic concepts compositions confuse the global-condition models, and the multimodal distribution hinders its capacity for controllable synthesis; 2) Removing the normalization in the condition vector has witnessed the realism degradation measured by FID, demonstrating its efficiency in reducing variance in latent space.

\begin{table}[H]
    \centering
    \small
    \begin{tabular}{cc|ccc}
    \toprule
    Training/Testing Dataset& Condition       &FID    &  KL  & CLAP\\
    \midrule
              AudioCaption  & Global          & /     &  /  &  0.12 \\
              FSD50k        & Global          &  40.7 & 8.2 &  0.40 \\  
              FSD50k        & NormGlobal      &  31.1 & 8.0 &  0.42 \\  
    \bottomrule            
    \end{tabular}
    \caption{Ablation studies for training Make-An-Audio with global conditions.}
    \label{table:ablation_v2a}
    \end{table}

\section{Dynamic Reprogramming Templates}   \label{templates}

Below we provide the list of text templates used when providing dynamic reprogramming:

\begin{itemize}
    \item before $v$ $q$ $a$ $n$ of \&, X
    \item X before $v$ $q$ $a$ $n$ of \&,
    \item in front of $v$ $q$ $a$ $n$ of \&, X
    \item first is X second is $q$ $a$ $n$ of \&
    \item after X, $v$ $q$ $a$ $n$ of \&
    \item after $v$ $q$ $a$ $n$ of \&, X
    \item behind $v$ $q$ $a$ $n$ of \&, X
    \item $v$ $q$ $a$ $n$ of \&, then X
    \item $v$ $q$ $a$ $n$ of \&, following X
    \item $v$ $q$ $a$ $n$ of \&, later X
    \item X after $v$ $q$ $a$ $n$ of \&
    \item before X, $v$ $q$ $a$ $n$ of \&
\end{itemize}

Specifically, we replace X and \&, respectively, with the natural language of sampled data and the class label of sampled events from the database.

For verb (denoted as v), we have \{`hearing', `noticing', `listening to', `appearing'\};
for adjective (denoted as a), we have \{`clear', `noisy', `close-up', `weird', `clean'\};
for noun (denoted as n), we have \{`audio', `sound', `voice'\};
for numeral/quantifier (denoted as q), we have \{`a', `the', `some'\};

\section{Potential Negative Societal Impacts} 

This paper aims to advance open-domain text-to-audio generation, which will ease the effort of short video and digital art creation. The efficient training method also transfers knowledge from text-to-audio models to X-to-audio generation, which helps avoid training from scratch, and thus reduces the issue of data scarcity. A negative impact is the risk of misinformation. To alleviate it, we can train an additional classifier to discriminate the fakes. We believe the benefits outweigh the downsides.

Make-An-Audio lowers the requirements for high-quality text-to-audio synthesis, which may cause unemployment for people with related occupations, such as sound engineers and radio hosts. In addition, there is the potential for harm from non-consensual voice cloning or the generation of fake media, and the voices in the recordings might be overused than they expect.

\section{Limitations}

Make-An-Audio adopts generative diffusion models for high-quality synthesis, and thus it inherently requires multiple iterative refinements for better results. Besides, latent diffusion models require typically require more computational resources, and degradation could be witnessed with decreased training data. One of our future directions is to develop lightweight and fast diffusion models for accelerating sampling.

\end{document}